\documentclass[12pt, a4paper]{iopart}
\usepackage{graphics,graphicx,epsfig}
\usepackage{iopams}

\begin{document}

\title{Prediction and typicality in multiverse cosmology}
\author{Feraz Azhar}
\address{Unit for History and Philosophy of Science, The University of Sydney, NSW 2006, Australia}
\ead{feraz.azhar@alumni.physics.ucsb.edu}

\begin{abstract}
In the absence of a fundamental theory that precisely predicts values for observable parameters, anthropic reasoning attempts to constrain probability distributions over those parameters in order to facilitate the extraction of testable predictions. The utility of this approach has been vigorously debated of late, particularly in light of theories that claim we live in a multiverse, where parameters may take differing values in regions lying outside our observable horizon. Within this cosmological framework, we investigate the efficacy of top-down anthropic reasoning based on the weak anthropic principle. We argue contrary to recent claims that it is not clear one can either dispense with notions of typicality altogether or presume typicality, in comparing resulting probability distributions with observations. We show in a concrete, top-down setting related to dark matter, that assumptions about typicality can dramatically affect predictions, thereby providing a guide to how errors in reasoning regarding typicality translate to errors in the assessment of predictive power. We conjecture that this dependence on typicality is an integral feature of anthropic reasoning in broader cosmological contexts, and argue in favour of the explicit inclusion of measures of typicality in schemes invoking anthropic reasoning, with a view to extracting predictions from multiverse scenarios.
\end{abstract}

\maketitle

\section{Introduction\label{SEC:Introduction}}

Generating testable predictions from theories positing the existence of a multiverse remains one of the key barriers to the broader appeal of multiverse proposals. There are a variety of contemporary theoretical ideas that give rise to multiverse proposals, including those based on inflationary cosmology~\cite{vilenkin_83, linde_83, linde_86}, and the string theory landscape~\cite{bousso+polchinski_00, kachru+al_03, freivogel+al_06, susskind_07}. In each case, short of a definite prediction for parameters of interest, one expects the theory to provide a probability distribution spread over some range, out of which parameters take their values. Comparisons between values that we observe and the values that are predicted by the theory through the distribution, can be used as a way of assessing the theory's predictive power, can help explain why parameters take their observed values, and can in principle help us determine whether the observed values provide support for the theory. This multifaceted role, coupling prediction, explanation and potential theory confirmation, makes the task of understanding how to compare observed values with a theory-generated probability distribution, an important step in establishing the validity of proposals for the multiverse. 

Owing to the presumed parsimony of description that any fundamental theory of the universe is likely to exhibit, together with the complex nature of our local conditions, a probability distribution generated from theory alone may not be enough to narrow down observations that we are likely to make~\cite{hartle_07}. It has been argued that conditionalization may be necessary to restrict the distribution in such a way that sharpens the comparison between theory and observation~\cite{hartle_07, aguirre+tegmark_05, weinstein_06}. There exist a range of approaches one can take, which in effect institute successively more restricted versions of the weak anthropic principle~\cite{carter_74}. The least restricted version does not conditionalize the distribution at all and has been referred to as the ``bottom-up" approach~\cite[p.~3]{aguirre+tegmark_05}. The most restricted version includes every piece of experimental information available, and is known as the ``top-down" approach~\cite[p.~3]{aguirre+tegmark_05}. An intermediate choice, which one can think of as interpolating between the two, is that of selecting a set of constraints that we believe are compatible with our existence, though not necessarily inclusive of every piece of available experimental evidence. This approach represents a way to restrict the bottom-up distribution to one that explicitly references our existence, while avoiding the complexity demanded by the top-down description, making the requisite comparison between our observations of a parameter of interest and the distribution out of which that parameter is drawn, potentially more immediate. This strategy has been termed the ``anthropic approach'', though we will refer to anything that is not the bottom-up approach and that lies on this ``spectrum of conditionalization'' as an anthropic approach (see Aguirre and Tegmark~\cite[p.~3]{aguirre+tegmark_05}).

The choice of a particular conditionalization scheme may well not be enough, however. A second choice needs to be made regarding whether one believes that our observations are \emph{randomly} selected from the distribution. Namely, are we a typical member of the class of observers implicitly defined by the conjunction of the theory and the conditionalization scheme~\cite{weinstein_06, hartle+srednicki_07, garriga+vilenkin_08, srednicki+hartle_10, bostrom_02}? Indeed, as we will examine, the predictions of this conjunction crucially depend upon the answer that one adopts to this question.\footnote{The assertion that we are typical, is commonly referred to as the ``principle of mediocrity", as formulated by Vilenkin~\cite[p.~847]{vilenkin_95}.}

The two controversies then, regarding a suitable conditionalization scheme for a theory-generated probability distribution, and whether we can then assume we are typical of the observers implied, have shaped recent debates into the predictive power of multiverse proposals. Weinstein advocates the top-down approach as the preferred method of conditionalization, and claims in that context, one can do away with the need for a stance on typicality with respect to the resulting distribution~\cite{weinstein_06}. Garriga and Vilenkin support prediction based upon the presumption that we are members of a reference class of observers who have ``identical information content"~\cite[p.~043526-1]{garriga+vilenkin_08}, in which case they claim that we can assume we are typical observers. Srednicki and Hartle argue that even within the context of conditionalization that rests on the presumption of different observers in the  multiverse having identical data, assumptions about typicality need to be built-in to the theory-generated distribution, and that predictive power should thereby be evaluated in the context of a conjunction consisting of the theory, the conditionalization scheme and assumptions regarding typicality~\cite{srednicki+hartle_10}. In a sense then, each of these three proposals operate within a scheme most similar to the top-down approach to conditionalization, but assume very different positions regarding how one should treat typicality, and therefore make radically different statements about how one should compare observations with theory.

In this paper we address this controversy from complementary points of view. We argue that conceptually, it is not clear who or what the constraints proposed in the context of top-down schemes imply, and that therefore any presumption of a lack of need to invoke a stance on typicality, or indeed the presumption of typicality itself is premature. We then show by way of a concrete example related to recent discussions about dark matter, that top-down conditionalization allows for mutually exclusive predictions regarding the number of dominant species contributing to the total dark matter density, depending upon the assumptions one makes about typicality. In this way, one's stance on typicality strongly influences the prediction, and then an incorrect presumption about the former can translate into a misunderstanding about the latter. We thereby side with Srednicki and Hartle in proposing that if one is interested in extracting predictions from theory-generated probability distributions via top-down reasoning, one must explicitly include an exhaustive set of typicality assumptions in the analysis~\cite{srednicki+hartle_10}. Without such an approach, anthropic reasoning may play little role in helping to extract predictions from multiverse proposals.

\section{Inherent ambiguities in uniquely characterizing `us'\label{SEC:The weak anthropic principle}}

There exist multiverse scenarios within the context of both inflationary cosmology and string theory that give rise to physical conditions in which our domain (or universe) is unlikely. Rather than consequently dismissing these scenarios by virtue of them not being predictive (of our domain), the weak anthropic principle (WAP) can be invoked to help refine the analysis of their predictive power. This principle was introduced by Carter in 1974, in the context of attempting to explain coincidences between cosmological parameters. It takes the following form:
\begin{quotation}
$\textrm{WAP}$: ``...what we can expect to observe must be restricted by the conditions necessary for our presence as observers."~\cite[p.~291]{carter_74}.
\end{quotation}
The idea is that one can use the WAP to restrict attention to the types of domains that we might expect to see, as generated by the theory under consideration, and then one can ask whether the restricted theory is compatible with our observations. It is clear that there is a large ambiguity inherent in the WAP, in that it doesn't specify precisely who or what it is referring to. Weinstein attempts to address this ambiguity, and in doing so claims that there exists a version of the WAP that does away with the need for subsequent presumptions about typicality in order to extract predictions~\cite{weinstein_06}. We argue that this is far from clear. 

Weinstein demarcates two versions of this principle. In his notation, we can interpret the WAP to mean either of the following:
\begin{quotation}
$\textrm{WAP}_{1}$: ``What we can expect to observe must be restricted by the conditions necessary for our presence."~\cite[p.~4234]{weinstein_06}, 
\end{quotation}
\begin{quotation}
$\textrm{WAP}_{2}$: ``What we can expect to observe must be restricted by the conditions necessary for the presence of observers."~\cite[p.~4234]{weinstein_06}.
\end{quotation}

The crux of Weinstein's argument is that if we think of the WAP as referring to conditions that are necessary for ``the presence of observers" (Weinstein's $\textrm{WAP}_{2}$), then we need to specify what these observers are, and to invoke a measure of (our) typicality within this reference class in order to extract a testable prediction. Indeed the term ``observers" invokes a potentially large class of beings of which we may or may not be  typical. Weinstein's issue with this approach is precisely the need to invoke the (potentially) ``misguided and unjustified" assumption of typicality~\cite[p.~4231]{weinstein_06}.

If, however, we  think of the WAP as referring to conditions that are necessary for \emph{us} (Weinstein's $\textrm{WAP}_{1}$), Weinstein argues that the probability distributions that obtain have immediate significance for the predictive power of the theory under consideration. If that theory predicts that \emph{we} should observe the values that we do with a non-zero probability, then it is in fact correctly predicting our observation. The purported lack of reliance of this version of the $\textrm{WAP}$ on (what he deems to be)  the contentious extra assumption of typicality, leads him to endorse the use of $\textrm{WAP}_{1}$ over $\textrm{WAP}_{2}$. 

Our concern is that it is not clear how to characterize `us', i.e., what is it that we are supposed to be characterizing and how does one translate this into a set of physical constraints that imply the existence of the thing being characterized in an unambiguous way? Ambiguities in addressing this concern will translate into ambiguities in interpreting the predictive power of any probability distribution that falls out of the theory. In light of this uncertainty, it appears that a notion of typicality might need to be included in any analysis resting upon $\textrm{WAP}_{1}$, and then Weinstein's misgivings about $\textrm{WAP}_{2}$ carry over to $\textrm{WAP}_{1}$ as well. 
 
Weinstein offers a way to partially address our concern by suggesting that we condition on ``as \emph{detailed} a description as possible"~\cite[p.~4235]{weinstein_06}, namely, that we should condition on observed values of cosmological parameters (say) in order to make subsequent predictions. This suggestion mirrors one made by Garriga and Vilenkin~\cite{garriga+vilenkin_08}. They argue that the reference class of observers who share ``identical information content" with us is a class within which we can assume typicality~\cite[p.~043526-1]{garriga+vilenkin_08}. Our concern in this context is that if we condition along the lines of either account, it is not clear who or what is implied by the conjunction of the underlying theory and the constraints. We cannot deduce what `everything we know' or the entirety of our information content gives rise to, and so it may be premature to ignore issues regarding typicality as suggested in the former case, or indeed to presume typicality as recommended in the latter. In the context of Garriga and Vilenkin's suggestion, our concern regarding the presumption of typicality manifests with greater immediacy given their own admission that from an operational point of view, it would be difficult to include in the analysis ``the full information content of observers" and that ``by necessity, we need to consider a reference class of observers specified by a small subset of all available information"~\cite[p.~043526-2]{garriga+vilenkin_08}.  It seems that only approximate conditions that specify ``our presence" can be invoked in any practical calculation. The basic indeterminacy that accompanies any attempt to narrow down a class of observers who accord in an unambiguous way with us, makes it difficult to see how one can suspend the need to take a stance regarding typicality (as Weinstein advocates), or indeed how one can assert it without worrying about the potential for generating spurious predictions (as is potentially the case with the argument of Garriga and Vilenkin). 

In the following section we take a more quantitative viewpoint and show that in at least one (rather stylized) theoretical example, conditionalizing one's distribution in a way that accords with the top-down approach gives results that clearly depend upon whether one assumes typicality with respect to the reference class implicitly invoked. We characterize how this dependence manifests itself, providing a sense of the nature of the predictive errors one is vulnerable to, in the case where one takes an incorrect stance on typicality. 

\section{Typicality and dark matter\label{SEC:Typicality and dark matter}}

One cosmological context in which arguments of the above form can play an important role is that of dark matter. The most popular current theory concerning the composition of dark matter is that it exists in the form of new non-baryonic particle species~\cite{strigari_12, bertone+al_05}. There are a broad range of candidate particles that have been put forward, including weakly interacting massive particles, for which vigorous experimental searches are currently in progress~\cite{agnese+al_13}. It has not  been ruled out that more than one dark matter particle candidate could contribute to the total dark matter density of the universe~\cite{aguirre+tegmark_05, bertone+al_05, tegmark+al_06}, and it is within this context that we will focus our theoretical concerns.

Consider then the case where there are $N > 1$ dark matter particle species that contribute to the observed dark matter density. Following~\cite{aguirre+tegmark_05}, we assume that these species have densities $\{\rho_{i}\}_{i = 1}^{N}$, which are given by a dimensionless dark matter-to-baryon ratio, $\rho_{i}:= \Omega_{i}/\Omega_{\textrm{\scriptsize{b}}}$, for $i = 1, \dots, N$. These densities represent a random variable whose joint probability distribution is set by a theory that gives rise to a multiverse, with dark matter density components sampled from this distribution. We are interested in attempting to predict the total number of dark matter particle species that contribute to the observed dark matter density $\rho_{\textrm{\scriptsize{o}}} = \sum_{i = 1}^{N}\rho_{i}$, while closely monitoring how this prediction varies with typicality assumptions.  We are furthermore interested in the top-down case of Aguirre and Tegmark~\cite{aguirre+tegmark_05}, which corresponds most closely to Weinstein's $\textrm{WAP}_{1}$ (see also Aguirre~\cite{aguirre_07}). In setting up the problem (section~\ref{SUBSEC:AT_calculation}), we closely follow the exposition of~\cite{aguirre+tegmark_05}. 

\subsection{The top-down approach of Aguirre and Tegmark}\label{SUBSEC:AT_calculation}

We assume that the dark matter particle species are independent, i.e., the joint probability density function $P(\rho_{1}, \rho_{2}, \dots, \rho_{N})$ factorizes into a product of marginals,
\begin{equation}
P(\rho_{1}, \rho_{2}, \dots, \rho_{N}) = \prod_{i= 1}^{N}P_{i}(\rho_{i}).
\end{equation}
We will consider those components that have significant probability of occurrence near or above the observed value $\rho_{i}\sim\rho_{\textrm{\scriptsize{o}}}$ as (following~\cite{aguirre+tegmark_05}) we'll assume that those components with significant probability weight at densities much smaller than $\rho_{\textrm{\scriptsize{o}}}$, do in fact take those values, since there is no contradiction with this assumption and the constraint that $\sum_{i=1}^{N}\rho_{i} = \rho_{\textrm{\scriptsize{o}}}$. This leaves $N_{\textrm{\scriptsize{big}}}$ dark matter components. We furthermore work with the differential distribution in $\ln(\rho_{i})$, that is, $\mathcal{P}_{i}:= \textrm{d}p_{i}(\rho_{i})/\textrm{d}\ln\rho_{i} = \rho_{i}P_{i}(\rho_{i})$, where $p_{i}({\rho_{i}})$ is the cumulative probability distribution function for the density of species $i$. 

Consider then the case of $N_{\textrm{\scriptsize{big}}} > 1$.\footnote{In what follows, we'll assume we've relabelled the indices over the dark matter densities such that the first $N_{\textrm{\scriptsize{big}}}$ indices refer to the $N_{\textrm{\scriptsize{big}}}$ significant dark matter components.}  First, we'd like to find the densities of the  $N_{\textrm{\scriptsize{big}}}$ components that maximize the total probability $\mathcal{P}_{\textrm{\scriptsize{tot}}} \propto \prod_{j}\mathcal{P}_{j}$, subject to the fact that we observe a total dark matter density of $\rho_{\textrm{\scriptsize{o}}} = \sum_{j}\rho_{j}$. This is simply a constrained optimization problem whose solution can be found by using a Lagrange multiplier $\lambda$, as in~\cite{aguirre+tegmark_05}. We want to maximize the expression $\mathcal{P}_{\textrm{\scriptsize{tot}}} - \lambda\sum_{j}\rho_{j}$, or equivalently, $\ln\mathcal{P}_{\textrm{\scriptsize{tot}}} - \lambda\sum_{j}\rho_{j}$. Noting that the distribution $\mathcal{P}_{\textrm{\scriptsize{tot}}}$ factorizes, the solution is $\textrm{d}\ln\mathcal{P}_{i}/\textrm{d}\rho_{i} = \lambda$~\cite{aguirre+tegmark_05}. One can then fix the Lagrange multiplier $\lambda$ by enforcing the constraint $\rho_{\textrm{\scriptsize{o}}} = \sum_{j}\rho_{j}$. Assuming a power law relation for the probabilities, 
\begin{equation}\label{EQN:TD_powerlaw}
\mathcal{P}_{i} \propto \rho_{i}^{\beta_{i}},
\end{equation}
one can readily show that the solution to the constrained optimization problem is given by 
\begin{equation}\label{EQN:TD_maxdensity}
\rho_{i} = \left(\frac{\beta_{i}}{\beta}\right) \rho_{\textrm{\scriptsize{o}}},
\end{equation}
where $\beta = \sum_{i}\beta_{i}$. Aguirre and Tegmark go on to claim that although the exponents $\beta_{i}$ may be different, rendering many fewer dominant dark matter components than $N_{\textrm{\scriptsize{big}}}$, in the absence of any further information, there is an argument that can be made that they are in fact similar, and that thereby there could be many components that contribute to the total dark matter density. 

In this fashion, they conclude that under top-down conditionalization, that is, where one conditionalizes the distribution based on one's observations (in this case, the total dark matter density), one predicts the existence of several dark matter species, contributing roughly equally to the total dark matter density. 

It is important to note that in the above argument, Aguirre and Tegmark have assumed that the condition that maximizes the probability is the one that determines what the theory predicts. In other words, they have assumed a very specific form of typicality in their construction, which presumes we will observe values that are most likely to occur, given the theory and the conditionalization scheme adopted. In what follows, we look into the consequences of relaxing this assumption, and show under this condition that a radically different picture emerges. 

\subsection{Relaxing typicality}

Consider then what happens to the predictions of section~\ref{SUBSEC:AT_calculation} when we do not demand optimality from our distribution. We'll first consider the simplest nontrivial case where the number of significant dark matter particle species contributing to the total dark matter density is $N_{\textrm{\scriptsize{big}}} = 2$. 

\subsubsection{$\boldsymbol{N_{\textrm{\scriptsize{{\bf big}}}} = 2}$.}

In this case, the distribution $\mathcal{P}_{\textrm{\scriptsize{tot}}}$ takes on its maximal value, $\mathcal{P}_{\textrm{\scriptsize{tot}}}^{\textrm{\scriptsize{MAX}}}$, when equations~(\ref{EQN:TD_powerlaw}) and (\ref{EQN:TD_maxdensity}) hold with $i = 1,2$:
\begin{eqnarray}
\mathcal{P}_{\textrm{\scriptsize{tot}}}^{\textrm{\scriptsize{MAX}}} &\propto& \left[\left(\frac{\beta_{1}}{\beta}\right) \rho_{\textrm{\scriptsize{o}}}\right]^{{\beta_1}}\left[\left(\frac{\beta_{2}}{\beta}\right) \rho_{\textrm{\scriptsize{o}}}\right]^{{\beta_2}} \nonumber \\
& = &  \rho_{\textrm{\scriptsize{o}}}^{\beta_{1}+\beta_{2}}\left(\frac{\beta_{1}}{\beta}\right)^{{\beta_1}}\left(\frac{\beta_{2}}{\beta}\right)^{{\beta_2}}.
\end{eqnarray}
The prediction in section~\ref{SUBSEC:AT_calculation}, of multiple dark matter species contributing to the total dark matter density depends on $\beta_{1}\approx\beta_{2}$. We'll investigate the suboptimal situation under the assumption that the power laws determining the behaviour of the relevant distributions have the same exponent, where the value of that exponent is consistent with the requirement that we're looking at two components with significant probability near the relatively high value of $\rho_{\textrm{\scriptsize{o}}}$. To that end, let $\beta_{1} = \beta_{2} = \beta^{\star} > 1$. The optimal $\rho_{i}^{\textrm{\scriptsize{MAX}}}$ then becomes $\rho_{i}^{\textrm{\scriptsize{MAX}}} = (1/2)\rho_{\textrm{\scriptsize{o}}}$. Also, 
\begin{equation}
\mathcal{P}_{\textrm{\scriptsize{tot}}}^{\textrm{\scriptsize{MAX}}} \propto \rho_{\textrm{\scriptsize{o}}}^{2\beta^{\star}}\left(\frac{1}{2}\right)^{2\beta^{\star}}.
\end{equation}

Now away from the maximum, on the constraint surface ($\rho_{1} +\rho_{2} = \rho_{\textrm{\scriptsize{o}}}$), 
\begin{equation}
\mathcal{P}_{\textrm{\scriptsize{tot}}} \propto \rho_{1}^{\beta^{\star}}(\rho_{\textrm{\scriptsize{o}}}-\rho_{1})^{\beta^{\star}}. 
\end{equation}
We'll consider the size of $\mathcal{P}_{\textrm{\scriptsize{tot}}}$ relative to $\mathcal{P}_{\textrm{\scriptsize{tot}}}^{\textrm{\scriptsize{MAX}}}$ as we scale $\rho_{1}$ away from its optimal value of $\rho_{1}^{\textrm{\scriptsize{MAX}}} = (1/2)\rho_{\textrm{\scriptsize{o}}}$. To this end let
\begin{equation}
\rho_{1} = \epsilon \frac{1}{2}\rho_{\textrm{\scriptsize{o}}}, 
\end{equation}
where  $0\leq\epsilon\leq 2$, and consider the resulting fractional probability, $\mathcal{P}_{\textrm{\scriptsize{tot}}}/\mathcal{P}_{\textrm{\scriptsize{tot}}}^{\textrm{\scriptsize{MAX}}}$. A quick calculation shows that 
\begin{equation}\label{EQN:N2_fracprob}
\frac{\mathcal{P}_{\textrm{\scriptsize{tot}}}}{\mathcal{P}_{\textrm{\scriptsize{tot}}}^{\textrm{\scriptsize{MAX}}}} = \left[\epsilon\left(2-\epsilon\right)\right]^{\beta^{\star}}. 
\end{equation}

The limit of interest here, can more clearly be explored by noting that equation~(\ref{EQN:N2_fracprob}) implies:
\begin{equation}\label{EQN:N2_epilson}
\epsilon = 1\pm \sqrt{1-(\mathcal{P}_{\textrm{\scriptsize{tot}}}/\mathcal{P}_{\textrm{\scriptsize{tot}}}^{\textrm{\scriptsize{MAX}}})^{1/\beta^{\star}}}.
\end{equation}
For a fixed $\beta^{\star}$, a sufficiently improbable situation, where $\mathcal{P}_{\textrm{\scriptsize{tot}}}/\mathcal{P}_{\textrm{\scriptsize{tot}}}^{\textrm{\scriptsize{MAX}}} \ll 1$, will force $\epsilon$ to take values near either 0 or 2. In either case, we are left with a single dominant dark matter species.\footnote{Our considerations are made under the assumption that the power law dependence of the probability (equation~(\ref{EQN:TD_powerlaw})) remains valid. A less stylized calculation would need to address the issue of potential deviations from this dependence when assessing how dominant one species could be in principle.}

Sufficiently far away from optimality then, the conclusion of the existence of multiple dark matter species contributing equally to the total observed dark matter density breaks down, and the prediction takes a new form. What we can expect to observe, depends upon whether we assume we will observe typical values from the distribution. Though perhaps this is to be expected, our point is that this holds even in the case of top-down conditionalization, namely, that assumptions regarding typicality play a role in the case most similar in spirit to the style of reasoning endorsed by Weinstein's $\textrm{WAP}_{1}$. For the sake of completeness, we'll show that this result also holds in the case where we let $N_{\textrm{\scriptsize{big}}} > 2$. 

\subsubsection{$\boldsymbol{N_{\textrm{\scriptsize{{\bf big}}}} = \mathcal{N} \geq 2}$.}

In the general case $N_{\textrm{\scriptsize{big}}} = \mathcal{N} \geq 2$, combining equations~(\ref{EQN:TD_powerlaw}) and (\ref{EQN:TD_maxdensity}) gives, 
\begin{equation}
\frac{\mathcal{P}_{\textrm{\scriptsize{tot}}}}{\mathcal{P}_{\textrm{\scriptsize{tot}}}^{\textrm{\scriptsize{MAX}}}} = \frac{1}{\rho_{\textrm{\scriptsize{o}}}^{\beta}}\prod_{i = 1}^{\mathcal{N}}\left(\frac{\beta}{\beta_{i}}\right)^{\beta_{i}} \rho_{i}^{\beta_{i}}.
\end{equation}
We'll again perform calculations in the case where the power law exponents have the same value, i.e., $\beta_{i} = \beta^{\star} >1$, for $i = 1,2, \dots, \mathcal{N}$. Noting then that $\beta = \mathcal{N}\beta^{\star}$, 
\begin{equation}
\frac{\mathcal{P}_{\textrm{\scriptsize{tot}}}}{\mathcal{P}_{\textrm{\scriptsize{tot}}}^{\textrm{\scriptsize{MAX}}}} = \left(\frac{\mathcal{N}}{\rho_{\textrm{\scriptsize{o}}}}\right)^{\mathcal{N}\beta^{\star}}\prod_{i= 1}^{\mathcal{N}}\rho_{i}^{\beta^{\star}}. 
\end{equation}

To investigate the behaviour of this function, we'll explore the possibility of $\rho_{i}$ being the dominant contributing species for $i=1$, though the argument works for any species $i$. In particular, we'll monitor the density $\rho_{1}$ as it deviates from its optimal value of $\rho_{1}^{\textrm{\scriptsize{MAX}}} = (1/\mathcal{N})\rho_{\textrm{\scriptsize{o}}}$, where this deviation is controlled by a parameter $\epsilon$:
\begin{eqnarray}
\rho_{1} &=& \epsilon\frac{1}{\mathcal{N}}\rho_{\textrm{\scriptsize{o}}}\label{EQN:rho1_generalN}, \\
\rho_{j}  &=& \rho_{\textrm{\scriptsize{o}}}\left(1-\frac{\epsilon}{\mathcal{N}}\right)\left(\frac{1}{\mathcal{N}-1}\right)\label{EQN:rhoj_generalN},
\end{eqnarray}
 with $0\leq\epsilon\leq\mathcal{N}$ and $j = 2, 3, \dots, \mathcal{N}$. We're thereby exploring the possibility of $\rho_{1}$ dominating with each of the other components being reduced by an equivalent amount with respect to their optimal values, such that the constraint $\sum_{i=1}^{\mathcal{N}}\rho_{i} = \rho_{\textrm{\scriptsize{o}}}$ is satisfied. 
 
\begin{figure}
\centering
\includegraphics{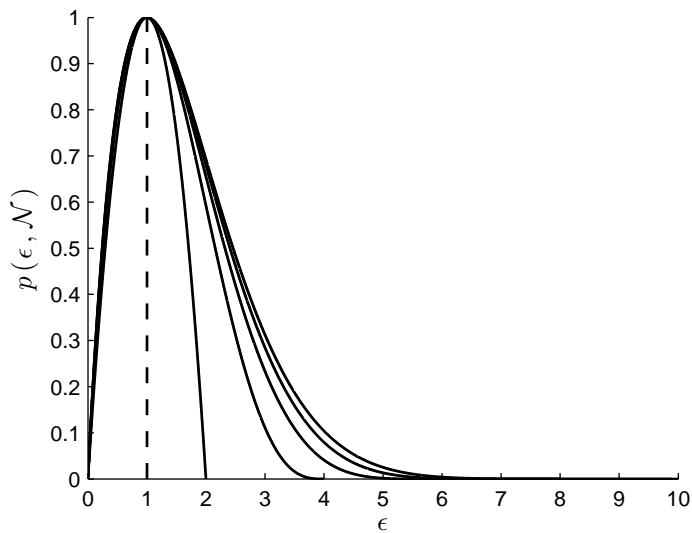}
\caption{Behaviour of $p(\epsilon, \mathcal{N})=\epsilon (\mathcal{N}-\epsilon)^{\mathcal{N}-1}/(\mathcal{N}-1)^{\mathcal{N}-1}$ for $N_{\textrm{\scriptsize{big}}} = \mathcal{N}$. The cases for $\mathcal{N} = 2,4,6,8,10$ are shown as a function of $\epsilon\in [0,\mathcal{N}]$. To the right of the peak, higher $\mathcal{N}$ cases taper to 0 for higher values of $\epsilon$. We note that for small values of $p(\epsilon, \mathcal{N})$, one can make the dark matter species represented by $\rho_{1}$ dominate over the other species, i.e, $\epsilon\to\mathcal{N}$. The dashed line ($\epsilon = 1$) corresponds to the optimal case of all $\mathcal{N}$ species contributing equally to the observed dark matter density $\rho_{\textrm{\scriptsize{o}}}$.
\label{FIG:F_brackets}}
\end{figure}

One can show then that the fractional probability (appropriately generalizing equation~(\ref{EQN:N2_fracprob})) is given by 
\begin{equation}\label{EQN:generalN_F}
\frac{\mathcal{P}_{\textrm{\scriptsize{tot}}}}{\mathcal{P}_{\textrm{\scriptsize{tot}}}^{\textrm{\scriptsize{MAX}}}} = \left[\frac{\epsilon (\mathcal{N}-\epsilon)^{\mathcal{N}-1}}{(\mathcal{N}-1)^{\mathcal{N}-1}}\right]^{\beta^{\star}}.
\end{equation}
The behaviour of the term in brackets in equation~(\ref{EQN:generalN_F}), which we'll refer to as $p(\epsilon, \mathcal{N}) := \epsilon (\mathcal{N}-\epsilon)^{\mathcal{N}-1}/(\mathcal{N}-1)^{\mathcal{N}-1}$,  is plotted in figure~\ref{FIG:F_brackets}. We see there that for appropriately low values of the fractional probability (depending of course upon $\beta^{\star}$), one can make the dark matter species with density $\rho_{1}$ dominate over the others. As $\mathcal{N}$ gets larger, $\rho_{1}$ dominates by a relatively smaller amount when compared with lower $\mathcal{N}$ values. For example, if $(\mathcal{P}_{\textrm{\scriptsize{tot}}}/\mathcal{P}_{\textrm{\scriptsize{tot}}}^{\textrm{\scriptsize{MAX}}})^{1/\beta^{\star}}= 0.010$, then $p(\epsilon, 2) = 0.010 \Longrightarrow \epsilon = 1.995$ (focusing only on the solution where $\rho_{1}$ dominates), giving $\rho_{1} = (1.995/2)\rho_{\textrm{\scriptsize{o}}}$ . On the other hand, $p(\epsilon, 10) = 0.010 \Longrightarrow \epsilon = 5.539$, giving $\rho_{1} = (5.539/10)\rho_{\textrm{\scriptsize{o}}}$. 

The upshot is that for a fixed $\beta^{\star}$, the nature of our prediction, represented here by the dominance of $\rho_{1}$, is controlled by how small we allow $(\mathcal{P}_{\textrm{\scriptsize{tot}}}/\mathcal{P}_{\textrm{\scriptsize{tot}}}^{\textrm{\scriptsize{MAX}}})^{1/\beta^{\star}}$ to be, namely, the degree of atypicality we are willing to accept. 

\section{The spectrum of typicality\label{SEC:The spectrum of typicality}}

Accompanying the conditionalization scheme one might adopt to generate predictions from a theory, is the decision one needs to make regarding the typicality of our observations. In section~\ref{SEC:Typicality and dark matter}, we demonstrated in a (single) concrete example, the dependence of predictions on typicality under the assumption of top-down conditionalization. One may argue that we have not gone far enough. That simply constraining one variable of interest (in our case the total dark matter density) isn't enough to uniquely specify (or even approximate) ``our presence" (as in $\textrm{WAP}_{1}$) and that thereby, one might expect to have to take a stance on how typical our observations are. We do not exclude this possibility, but we question whether the situation will change as we successively add further observational constraints (in the above top-down fashion) until we have a distribution for some observational parameter that is the result of conditionalizing a theory on ``precise values for all cosmological parameters" ~(Weinstein~\cite[p.~4235]{weinstein_06}). Weinstein adds that under these circumstances, ``there is no need at all for a principle of mediocrity -- no need to assume that we are typical members of some larger ensemble.  Taking selection effects as seriously as possible is thus equivalent to appealing to $\textrm{WAP}_{1}$  as a principle of inference."~\cite[p.~4235]{weinstein_06}. We contend that given a theory that proposes the existence of multiple copies of everything \emph{we} know, it is not clear that we can dispense with having to deal with whether we are typical of what that theory gives rise to. In fact (in contrast to Garriga and Vilenkin), we may well not be typical and then assuming that we are, will produce incorrect results. 

To restate these claims in more familiar terminology, we are interested in using our underlying theory $\mathcal{T}$ to make predictions for a yet unobserved parameter, say $\Lambda_{\textrm{\scriptsize{new}}}$.\footnote{Note that we only consider here the case where we are interested in the predictive power of some specified theory $\mathcal{T}$. We aren't necessarily comparing theories, in which case one might want to employ Bayesian analyses in the style of Hartle and Srednicki~\cite{hartle+srednicki_07, srednicki+hartle_10}.} We conditionalize the theory-generated probability distribution $P(\Lambda_{\textrm{\scriptsize{new}}}|\mathcal{T})$, by fixing the values of all cosmological parameters in the theory (say $\vec{\lambda}$) to their currently observed values (giving $\vec{\lambda}_{\textrm{\scriptsize{obs}}}$). Then from an operational point of view, Weinstein's claim amounts to computing $P(\Lambda_{\textrm{\scriptsize{new}}}|\mathcal{T}, \vec{\lambda}_{\textrm{\scriptsize{obs}}})$, and then using this distribution to predict our future measurement of $\Lambda_{\textrm{\scriptsize{new}}}$, without concerning ourselves with whether we are typical of the observers thus generated. In fact, he mentions that ``if the theory assigns a non-vanishing probability to the parameter values we do observe, then that is what \emph{we} should \emph{expect} to observe." ~\cite[p.~4235]{weinstein_06}. Our contention is that it is not clear precisely who or what is specified by fixing $\vec{\lambda}$ to our observed values, and that thereby one may be subject to the same ``inductive overreach" that Weinstein levels at $\textrm{WAP}_{2}$~\cite[p.~4231]{weinstein_06}.
 
In this sense, our argument aligns with the claims of Srednicki and Hartle~\cite{srednicki+hartle_10}, who argue (in the context of distributions conditionalized on our data) for the introduction of a \emph{xerographic distribution} that can enforce a spectrum of typicality assumptions. In fact they argue for exploring schemes in which one can vary both the underlying theory $\mathcal{T}$ and the xerographic distribution, to explore how predictive the combination of the two are. A successful prediction (of some future observation) is one where the distribution conditionalized on the conjunction of the theory, our data, and typicality assumption is peaked around the value that we subsequently observe. We contend that any calculation of the sort that gives rise to a distribution such as $P(\Lambda_{\textrm{\scriptsize{new}}}|\mathcal{T}, \vec{\lambda}_{\textrm{\scriptsize{obs}}})$ should indeed be supplemented by such a scheme in order to extract predictions. Simply conditionalizing on everything we know isn't enough. 

\section{Discussion\label{SEC:Discussion}}

It is difficult to see how one would be able to dispense with typicality assumptions when conditionalizing in a top-down fashion given that it is far from clear how to characterize \emph{us}. Weinstein's effective claim about not requiring assumptions of typicality in the context of distributions such as $P(\Lambda_{\textrm{\scriptsize{new}}}|\mathcal{T}, \vec{\lambda}_{\textrm{\scriptsize{obs}}})$ (from section~\ref{SEC:The spectrum of typicality}) requires that the sum total of observational evidence we have at our disposal determines the existence of creatures who accord in some unambiguous way with us. Garriga and Vilenkin's claim about the legitimacy of presuming typicality with respect to reference classes that share all of our information content (and a small subset of this information content in more concrete settings) is subject to the same concern regarding our ignorance about who or what this prescription implies. 

This problem becomes more acute if as we have shown in the restricted example of section~\ref{SEC:Typicality and dark matter}, the typical members of the distribution see one thing while the atypical members see another. In the particular context considered there, demanding a low fractional probability, i.e., demanding parameters sample from an atypical region of the probability distribution, leads (under appropriate constraints) to the possibility of a single dominant dark matter species. This prediction differs from the case where we accept only the most probable (or under the appropriate conditions, typical) dark matter densities as those that set the prediction.

In a sense, our argument here is in accord with what one might expect intuitively. Abstracting away from more likely scenarios isn't a priori inconsistent with obtaining a different result. The key point is that when it is not clear that top-down conditionalization leads uniquely to \emph{us}, and when prediction is intimately entwined with presumptions about typicality, either dispensing entirely with the need to consider typicality, or automatically assuming typicality, seems presumptuous at best, or leads to inferences that are plainly incorrect under less favourable circumstances. Typicality, then, plays an important role in interpreting results (see~\cite{maor+al_08}), and it may be best to explicitly include this in any analysis of predictive power as proposed by Srednicki and Hartle~\cite{srednicki+hartle_10}. 

Our focus has been on top-down reasoning, though we see no reason why our arguments about the importance of considering typicality do not carry over to other cases in Aguirre and Tegmark's ``spectrum of conditionalization"~\cite{aguirre+tegmark_05}. In this way, interactions between the twin spectra of conditionalization and typicality determine the predictive power of multiverse cosmological models, an interaction that needs to be accounted for when we propose to test these models' predictions against observational evidence. 

\ack 

I thank Dean Rickles for many helpful discussions.

\end{document}